# Estimation of Matusita Overlapping Coefficient $\rho$ for Pair Normal Distributions

Omar M. Eidous     and     Salam K. Daradkeh

omarm@yu.edu.jo          2018107008@ses.yu.edu.jo

Department of Statistics - Faculty of Science - Yarmouk University

Irbid - Jordan

## Abstract

The Matusita overlapping coefficient $\rho$ is defined as agreement or similarity between two or more distributions. The parametric normal distribution is one of the most important statistical distributions. Under the assumption that the data at hand follow two independent normal distributions, this paper suggests a new technique to estimate the Matusita coefficient $\rho$. In contrast to the studies in the literature, the suggested technique requires no assumptions on the location and scale parameters of the normal distributions. The finite properties of the resulting estimators are investigated and compared with the nonparametric kernel estimators and with some existing estimators via simulation techniques. The results show that the performance of the proposed estimators is better than the kernel estimators for all considered cases.

## 1. Introduction

There are three overlapping coefficients, namely; Matusita (1955) coefficient $\rho$, Morisita (1959) coefficient $\lambda$ and Weitzman (1970) coefficient $\Delta$. These coefficients represent the similarity or agreement between two or more probability distributions or populations represented by their distributions. This paper interests with the Matusita $\rho$, which is given by,

$$\rho = \int \sqrt{f_1(x) f_2(x)} dx$$

where $f_1(x)$ and $f_2(x)$ are two continuous probability density functions ($pdfs$) (see Mulekar and Mishra, 1994 and 2000). The values of this measure should be ranging from 0 to 1. Zero value indicates no common area between population densities, while one indicates the perfect agreement between population densities. Note that this measure can be applied directly on discrete distributions by replacing summation instead of integrals. The main problem here is to estimate the parameter $\rho$. In general, there are two methods used to estimate overlapping measures; the parametric and nonparametric methods. The parametric method requires an assumption that the form of the probability density functions should be known with unknown parameter $\Theta$, where $\Theta$ may be a vector of parameters estimated by traditional methods, like the method of moments (MM) or the maximum likelihood (ML) method. On the other hand, the nonparametric method does not require any assumptions about the functional form of the density function. As an example of the parametric method, Inman and Bradly (1989) derived the ML estimator of Weitzman coefficient $\Delta$ with the assumption the two densities are normal with different means and equal variances (see Figure 1). Mulekar and Mishra (1994) studied the overlapping measures of two normal densities in the case of equal means but different variances of the two densities (see

Figure 2). Reiser and Faraggi (1999) constructed generalized confidence intervals for the overlapping measures of two normal distributions with equal variances. To compare the confidence intervals of the overlapping measures, Mulekar and Mishra (2000) used some resampling techniques, namely, Jackknife, bootstrap and transformation methods under the assumption that the two densities are normal with equal means. Parametric methods for confidence interval estimation of $\Delta$ were studied by Wang and Tian (2017), who also proposed methods for confidence interval estimation of $\Delta$ under various distributions, including normal.

Madhuri et al. (2001) and Al-Saleh and Samawi (2007) estimated the overlapping measures of two exponential populations with different scale parameters. Al-Saidy et al. (2005) considered the overlapping measures for two Weibull distributions having the same shape parameter but different scale parameters. Samawi and Al-Saleh (2008) used the ranked set sampling method to draw inferences about the overlapping measures of two exponential distributions with different scale parameters. Moreover, they constructed the confidence intervals for the overlapping measures using bootstrap and Taylor series methods. Chaubey et al. (2008) addressed the point estimation of the three overlapping measures when the two populations are described by inverse Gaussian distributions with equal means. Helu and Samawi (2011) studied the effect of sampling procedures on the overlapping measures by comparing the simple random sample with the ranked set sampling when the original data follow the Lomax models.

In parametric methods mentioned above and for distributions with two parameters, the authors assumed that the two location parameters are equal, the two scale parameters are equal, or the two shape parameters are equal. This assumption is required to enable the authors to find a closed-form for the interested overlapping parameter aim to estimate. That is, to be able to find the integrals arise in formulas of the different overlapping measures.

If there are some doubts about the validity of the model assumption or if the model assumption is difficult to be satisfied then the nonparametric method can be used instead of the parametric method. Many authors have studied the nonparametric method to make inferences about overlapping measures coefficients. For example, see Clemons and Bradley (2000), Clemons (2001), Mizuno et al. (2005), Schmid and Schmidt (2006), Ridout and Linkie (2008), Martens et al. (2014) and Eidous and Al-Talafha (2020). Most of the above studies used the nonparametric kernel method to estimate the overlapping measure $\Delta$. Samawi et al. (2011) introduced the nonparametric kernel method to estimate the overlapping measures $\Delta$ in order to perform a test about the symmetry of the underlying distribution of the data. Alodat et al. (2021) studied the asymptotic properties and the asymptotic distribution of the kernel estimator of the overlapping measure $\rho$. They used the resulting estimator for the goodness of fit test for two independent distributions.

## 2. Matusita coefficient $\rho$ for two normal distributions

A normal distribution is one of the most common continuous probability distribution for real-valued random variable. The general form of its $pdf$ for a random variable $X$ is,

$$f_X(x) = \frac{1}{\sigma\sqrt{2\pi}} e^{-\frac{1}{2}\left(\frac{x-\mu}{\sigma}\right)^2}, -\infty < x < \infty,$$

which is denoted by $N(\mu, \sigma^2)$. The normal distribution is the most important probability distribution in statistics because it fits many natural phenomena. For example, heights, blood pressure, measurement error, and IQ scores follow the normal distribution. The two parameters $\mu$ and $\sigma$ represent the distribution mean and standard deviation respectively. For the normal distribution, moving the mean simply slides the curve left or right - it changes the center, not the spread. The standard deviation, on the other hand, changes the shape of distribution curve and show how it is spread around its mean. Illustrations of the two cases are shown in Figure (1) and Figure (2) respectively.

**2.1 Equal variances**

The Matusita coefficient $\rho$ between two normal distributions with equal variances ($\sigma_1^2 = \sigma_2^2 = \sigma^2 \ say$) is given by (Al-Daradkeh, 2021),

$$\rho = e^{-(\mu_1-\mu_2)^2/(8\sigma^2)}$$

Let $X_1, X_2, \ldots, X_{n_1}$ and $Y_1, Y_2, \ldots, Y_{n_2}$ be two independent random samples drawn from two normal densities $N(\mu_1, \sigma^2)$ and $N(\mu_2, \sigma^2)$ respectively. The maximum likelihood (ML) estimators of $\mu_1$, $\mu_2$ and $\sigma^2$ are $\bar{X}$, $\bar{Y}$ and $S^2 = \frac{\sum_{i=1}^{n_1}(X_i-\bar{X})^2 + \sum_{i=1}^{n_2}(Y_i-\bar{Y})^2}{n_1+n_2}$ respectively. Therefore, the ML estimator of $\rho$ is (Al-Daradkeh, 2021),

$$\hat{\rho}_1 = e^{-(\bar{X}-\bar{Y})^2/(8S^2)}.$$

**2.1 Equal means**

The Matusita coefficient $\rho$ between two normal distributions with equal means ($\mu_1 = \mu_2 = \mu \ say$) is,

$$\rho = \sqrt{\frac{2C}{1+C^2}}$$

where $C = \sigma_1/\sigma_2$ (see Mulekar and Mishra, 1994). Suppose that $X_1, X_2, \ldots, X_{n_1}$ and $Y_1, Y_2, \ldots, Y_{n_2}$ are two independent random samples drawn from two normal densities $N(\mu, \sigma_1^2)$ and $N(\mu, \sigma_2^2)$ respectively. Mulekar and Mishra (1994) gave the following estimators for $\mu$, $\sigma_1^2$ and $\sigma_2^2$,

$$\hat{\sigma}_1^2 = \frac{\sum_{i=1}^{n_1}(X_i - \hat{\mu})^2}{n_1}$$

$$\hat{\sigma}_2^2 = \frac{\sum_{i=1}^{n_2}(Y_i - \hat{\mu})^2}{n_2}$$

and

$$\hat{\mu} = \frac{\sum_{i=1}^{n_1} X_i + \sum_{i=1}^{n_2} Y_i}{n_1 + n_2}.$$

respectively. They gave the following estimator for $\rho$,

$$\hat{\rho}_2 = \sqrt{\frac{2\hat{C}}{1 + \hat{C}^2}}$$

where $\hat{C} = \hat{\sigma}_1/\hat{\sigma}_2$.

The general case that does not assume equality between location parameters and equality between scale parameters has not been addressed or studied in the literature. This paper addresses this case and proposes a new technique for estimating $\rho$ without using any assumptions about the pair normal distribution parameters.

### 3. Proposed method to estimate $\rho$

Let $X_1, X_2, \ldots, X_{n_1}$ and $Y_1, Y_2, \ldots, Y_{n_2}$ be two independent random samples, which are taken from two normal distributions $N(\mu_1, \sigma_1^2)$ and $N(\mu_2, \sigma_2^2)$ respectively. It is direct method to derive the ML estimators of $\mu_1, \mu_2, \sigma_1^2$ and $\sigma_2^2$, which are respectively given by

$$\hat{\mu}_1 = \frac{\sum_{i=1}^{n_1} X_i}{n_1} = \bar{X}$$

$$\hat{\mu}_2 = \frac{\sum_{i=1}^{n_2} Y_i}{n_2} = \bar{Y}$$

$$\hat{\sigma}_1^2 = \sum_{i=1}^{n_1}(X_i - \bar{X})^2 / n_1$$

and

$$\hat{\sigma}_2^2 = \sum_{i=1}^{n_2}(Y_i - \bar{Y})^2 / n_2.$$

Therefore, the ML estimators of $f_1(x) = N(\mu_1, \sigma_1^2)$ and $f_2(y) = N(\mu_2, \sigma_2^2)$ are $\hat{f}_1(x) = N(\hat{\mu}_1, \hat{\sigma}_1^2)$ and $\hat{f}_2(y) = N(\hat{\mu}_2, \hat{\sigma}_2^2)$, respectively.

To estimate $\rho$, we suggest to express $\rho$ as expected value (mean) of some functions as follows.

Let $\psi_1(x)$ be a continuous function of $x$ and $\psi_2(y)$ is another continuous function of $y$, then $E(\psi_1(X))^{1/2} = \int (\psi_1(x))^{1/2} f_1(x) dx$ and $E(\psi_2(Y))^{1/2} = \int (\psi_2(y))^{1/2} f_2(y) dy$. Now, by taking $\psi_1(x) = f_2(x)/f_1(x)$ and $\psi_2(y) = f_1(y)/f_2(y)$ then we obtain,

$$E\left(\frac{f_2(X)}{f_1(X)}\right)^{\frac{1}{2}} = \int_{-\infty}^{\infty} \left(\frac{f_2(x)}{f_1(x)}\right)^{\frac{1}{2}} f_1(x) dx$$

$$= \int_{-\infty}^{\infty} \sqrt{f_1(x) f_2(x)} \, dx$$

$$= \rho$$

and

$$E\left(\frac{f_1(Y)}{f_2(Y)}\right)^{\frac{1}{2}} = \int_{-\infty}^{\infty} \left(\frac{f_1(y)}{f_2(y)}\right)^{\frac{1}{2}} f_2(y) dy$$

$$= \int_{-\infty}^{\infty} \sqrt{f_1(y) f_2(y)} \, dy$$

$$= \int_{-\infty}^{\infty} \sqrt{f_1(x) f_2(x)} \, dx$$

$$= \rho$$

Based on the last two expressions of $\rho$, we can also write $\rho$ as the average of both expressions as follows,

$$\rho = \frac{1}{2}\left[E\left(\frac{f_2(X)}{f_1(X)}\right)^{\frac{1}{2}} + E\left(\frac{f_1(Y)}{f_2(Y)}\right)^{\frac{1}{2}}\right].$$

Now, by using the last three formulas of $\rho$ and based on the two independent random samples $X_1, X_2, \ldots, X_{n_1}$ and $Y_1, Y_2, \ldots, Y_{n_2}$, (where the first sample is taken from $N(\mu_1, \sigma_1^2)$ and the second one is taken from $N(\mu_2, \sigma_2^2)$), we suggest to estimate $\rho$ as follows,

Since $E(f_2(X)/f_1(X))^{\frac{1}{2}}$ is the mean of $(f_2(X)/f_1(X))^{\frac{1}{2}}$, we can estimate it based on the first sample by the mean of $\left(\widehat{f_2}(X_i)/\widehat{f_1}(X_i)\right)^{1/2}$, $i = 1, 2, \ldots, n_1$. That is,

$$\hat{\rho} = \frac{1}{n_1}\sum_{i=1}^{n_1}\left(\frac{\widehat{f_2}(X_i)}{\widehat{f_1}(X_i)}\right)^{1/2},$$

Similarly, $E(f_1(Y)/f_2(Y))^{\frac{1}{2}}$ can be estimated based on the second sample by the mean of $\left(\widehat{f_1}(Y_i)/\widehat{f_2}(Y_i)\right)^{1/2}$, $i = 1, 2, \ldots, n_2$. That is,

$$\hat{\rho} = \frac{1}{n_2}\sum_{i=1}^{n_2}\left(\frac{\widehat{f_1}(Y_i)}{\widehat{f_2}(Y_i)}\right)^{1/2}$$

and finally, $\rho = \frac{1}{2}\left[E\left(\frac{f_2(X)}{f_1(X)}\right)^{\frac{1}{2}} + E\left(\frac{f_1(Y)}{f_2(Y)}\right)^{\frac{1}{2}}\right]$ is estimated by,

$$\hat{\rho}_P = \frac{1}{2}\left[\frac{1}{n_1}\sum_{i=1}^{n_1}\left(\frac{\widehat{f_2}(X_i)}{\widehat{f_1}(X_i)}\right)^{1/2} + \frac{1}{n_2}\sum_{i=1}^{n_2}\left(\frac{\widehat{f_1}(Y_i)}{\widehat{f_2}(Y_i)}\right)^{1/2}\right].$$

where the symbol P in $\hat{\rho}_P$ stands for "Proposed". A preliminary simulation study showed that the last estimator of $\rho$ is more stable than the other two estimators. Therefore, the performance of the last estimator is investigated in our simulation study in the next section.

**4. Simulation Study and Results**

In this section, a simulation study was conducted to investigate the performances of the proposed estimator $\hat{\rho}_P$ of $\rho$. For sake of comparison, the two estimators $\hat{\rho}_1$ and $\hat{\rho}_2$, which are stated in Section (2) are also considered. The estimator $\hat{\rho}_1$ was derived under the assumption that the means of the pair normal distributions are equal, while the estimator $\hat{\rho}_2$ was developed under the assumption that their variances are equals. In addition, the nonparametric kernel estimator $\hat{\rho}_k$ (see Eidous and Al-Talafha, 2020) is also included in this study. The kernel estimator $\hat{\rho}_k$ requires no assumptions about the distributions parameters or even the functional form of the distribution itself (see Eidous and Al-Talafha, 2020). To study the performances of the above estimators, the two independent samples $x_1, x_2, \ldots, x_{n_1}$ and $y_1, y_2, \ldots, y_{n_2}$ were simulated from two independent normal distributions with specific values of the corresponding parameters. Based on the parameters selection, 9 pairs of normal distributions were simulated, 3 of them were taken to deal with the case of equal means. 3 pairs were taken for the case of equal variances. The remaining 3 pairs deal with the case of unequal means as well as the unequal variances. The values of the corresponding parameters were chosen arbitrary but to allow the overlapping measure to vary below 0.5, around 0.5 and above 0.5. The pairs of two normal distributions and the values of the selected parameters that used in this simulation study were presented in Table (1). To study the different estimators and their behavior for different sample sizes, we chose $(n_1, n_2) = (10,10), (20,30), (30,30)$ and $(100,200)$. For each sample size, R = 1000 samples were generated and the empirical measures, Relative Bias (RB) and Relative Mean Square Error (RMSE) were computed for each estimator. If $\hat{\eta}$ is the estimator of $\eta$ then,

$$RB = \frac{\hat{E}(\hat{\eta}) - \eta}{\eta},$$

and

$$RMSE = \frac{\sqrt{\widehat{MSE}(\hat{\eta})}}{\eta},$$

where

$$\hat{E}(\hat{\eta}) = \frac{\sum_{j=1}^{R} \hat{\eta}_{(j)}}{R}$$

and

$$\widehat{MSE}(\hat{\eta}) = \sum_{j=1}^{R} \left( \hat{\eta}_{(j)} - \hat{E}(\hat{\eta}) \right)^2 / R.$$

All simulation results are calculated by using Mathematica, Version 7.

Based on the simulation results, which presented in Table (2), Table (3) and Table (4), we can conclude the following:

1. It is clear that the $|RB|$s and RMSEs of the different estimators decrease as the sample sizes increases. This is a good sign to conclude that the different considered estimators are consistent estimators for $\rho$.

2. The $RB$ values associated the estimators $\hat{\rho}_k$ and $\hat{\rho}_P$ are negative in all considered cases, which indicates that the two estimators are underestimate the exact values of $\rho$.

3. As expected, the proposed estimator $\hat{\rho}_P$ performs better than the kernel estimator $\hat{\rho}_k$ in all considered cases and for all values of the samples size. This is clear when we examine the corresponding values of $RMSE$ for each estimator. It is worthwhile to mention here that the proposed estimator was derived under the pair normal distributions, while the kernel estimator was developed as a general nonparametric method without using any functional form for the data at hand (See Eidoos and Al-Talafha, 2020).

4. The performances of the two estimators $\hat{\rho}_1$ and $\hat{\rho}_2$ are better than the other estimators. However, the drawback of these two estimators is that we can use each of them only if the corresponding required assumption is valid. The equal means assumption is required for $\hat{\rho}_1$ and the assumption of equal variances is required for $\hat{\rho}_2$.

5. The $RMSE$ values associated with the different estimators increase as the exact value of $\rho$ decreases. However, it is not necessarily correct to say that the mean square error ($MSE$) of these estimators increases with decreasing $\rho$, because the $RMSE$ formula is the square root of $MSE$ but divided by the exact value of $\rho$.

## 5. Discussion

The objective of this paper was to develop a new technique to estimate the Matusita coefficient (measure) $\rho$ under pair normal distributions. The advantage of the new technique over existing methods is that it does not require any assumptions about the parameters of the distributions. In addition, the proposed technique in this paper can be used under different parametric distributions such as Weibull, beta, Lomax, etc. The numerical results showed the good performance of the resulting estimator based on the new proposed technique.

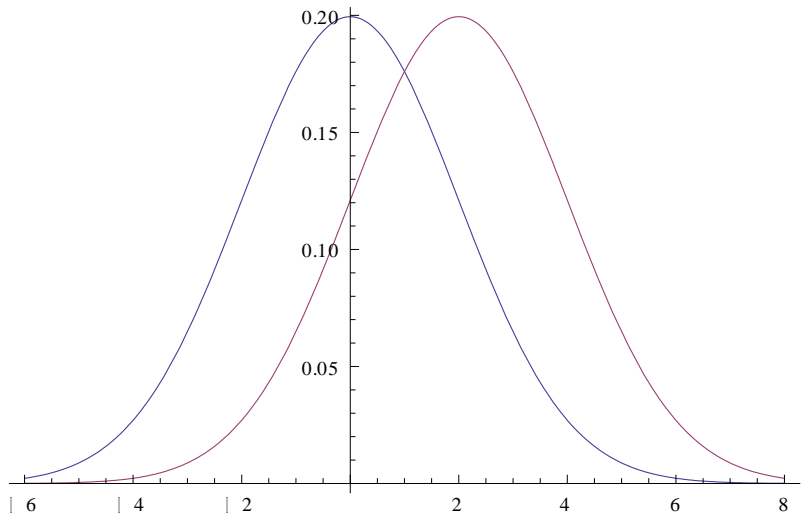

**Figure (1).** Two normal distributions with different means and equal variances.

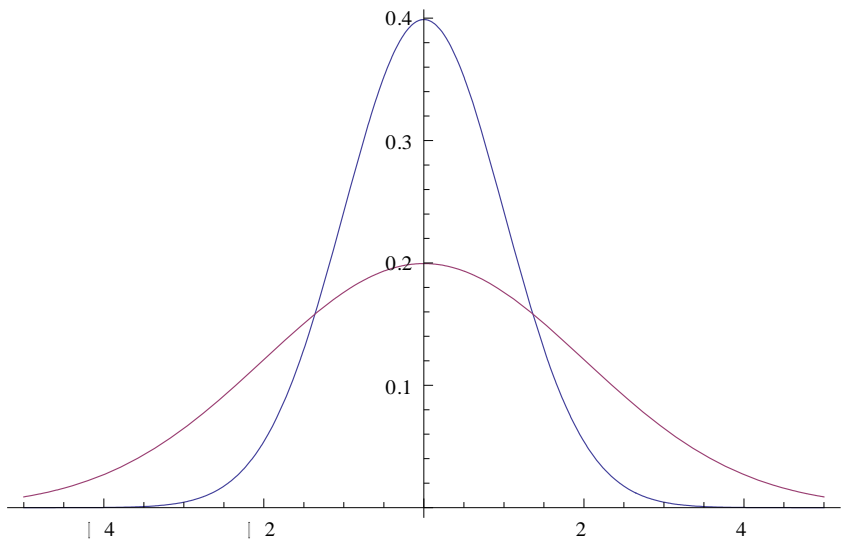

**Figure (2).** Two normal distributions with different variances and equal means.

**Table (1).** The 9 simulated pair normal distributions $f_X(x)$ and $f_Y(y)$.

|  | $f_X(x)$ | $f_Y(y)$ |
|---|---|---|
| Equal means | $N(0,1)$ | $N(0,1.5)$ |
|  | $N(0,1)$ | $N(0,2.5)$ |
|  | $N(0,1)$ | $N(0,10)$ |
| Equal variances | $N(0,1)$ | $N(-0.5,1)$ |
|  | $N(0,1)$ | $N(1.5,1)$ |
|  | $N(0,1)$ | $N(3,1)$ |
| Different means and different variances | $N(0,1)$ | $N(-0.2,1.1)$ |
|  | $N(0,1)$ | $N(2.5,4)$ |
|  | $N(0,1)$ | $N(5,2)$ |

**Table (2)**. The RB and RMSE of the estimators $\hat{\rho}_k, \hat{\rho}_l, \hat{\rho}_p$ when the data are simulated from pair normal distributions with equal means ($\mu_1 = \mu_2 = 0$).

| $(\sigma_1^2, \sigma_2^2)$ | $(n_1, n_2)$ | | $\hat{\rho}_k$ | $\hat{\rho}_l$ | $\hat{\rho}_p$ |
|---|---|---|---|---|---|
| colspan="5" Exact $\rho = 0.9607$ | | | | | |
| (1,1.5) | (10,10) | RB | -0.1197 | -0.0132 | -0.0292 |
| | | RMSE | 0.1616 | 0.0573 | 0.0788 |
| | (20,30) | RB | -0.0648 | -0.0075 | -0.0143 |
| | | RMSE | 0.0873 | 0.0404 | 0.0496 |
| | (30,30) | RB | -0.0525 | -0.0063 | -0.0114 |
| | | RMSE | 0.0715 | 0.0369 | 0.0438 |
| | (100,200) | RB | -0.0192 | -0.0014 | -0.0027 |
| | | RMSE | 0.0270 | 0.0156 | 0.0168 |
| colspan="5" Exact $\rho = 0.8304$ | | | | | |
| (1,2.5) | (10,10) | RB | -0.1192 | 0.0198 | -0.0307 |
| | | RMSE | 0.2041 | 0.1103 | 0.1472 |
| | (20,30) | RB | -0.0682 | 0.0045 | -0.0176 |
| | | RMSE | 0.1163 | 0.0723 | 0.0892 |
| | (30,30) | RB | -0.0572 | 0.0055 | -0.0130 |
| | | RMSE | 0.1023 | 0.0669 | 0.0836 |
| | (100,200) | RB | -0.0247 | 0.0019 | -0.0023 |
| | | RMSE | 0.0440 | 0.0314 | 0.0370 |
| colspan="5" Exact $\rho = 0.4449$ | | | | | |
| (1,10) | (10,10) | RB | -0.1275 | 0.2964 | -0.0446 |
| | | RMSE | 0.3698 | 0.4093 | 0.3353 |
| | (20,30) | RB | -0.0551 | 0.1747 | 0.0015 |
| | | RMSE | 0.2089 | 0.2659 | 0.1986 |
| | (30,30) | RB | -0.0585 | 0.1422 | -0.0113 |
| | | RMSE | 0.2047 | 0.2182 | 0.1930 |
| | (100,200) | RB | -0.0316 | 0.0467 | -0.0038 |
| | | RMSE | 0.0863 | 0.0859 | 0.0846 |

**Table(3).** The RB and RMSE of the estimators $\hat{\rho}_k, \hat{\rho}_l$ and $\hat{\rho}_p$ when the data are simulated from pair normal distributions with equal variances ($\sigma_1^2 = \sigma_2^2 = 1$).

| ($\mu_1, \mu_2$) | ($n_1, n_2$) | | $\hat{\rho}_k$ | $\hat{\rho}_2$ | $\hat{\rho}_p$ |
|---|---|---|---|---|---|
| colspan="5" | Exact $\rho = 0.9692$ | | | | |
| (0, -0.5) | (10,10) | RB | -0.1373 | 0.0307 | -0.0679 |
| | | RMSE | 0.1801 | 0.0826 | 0.1153 |
| | (20,30) | RB | -0.0653 | -0.0114 | -0.0250 |
| | | RMSE | 0.0863 | 0.0444 | 0.0540 |
| | (30,30) | RB | -0.0573 | -0.0107 | -0.0217 |
| | | RMSE | 0.0736 | 0.0372 | 0.0448 |
| | (100,200) | RB | -0.0173 | -0.0014 | -0.0037 |
| | | RMSE | 0.0247 | 0.0157 | 0.0167 |
| colspan="5" | Exact $\rho = 0.7548$ | | | | |
| (0, 1.5) | (10,10) | RB | -0.1660 | -0.03186 | -0.09260 |
| | | RMSE | 0.2801 | 0.1984 | 0.2305 |
| | (20,30) | RB | -0.0799 | -0.0046 | -0.0306 |
| | | RMSE | 0.1586 | 0.1236 | 0.1346 |
| | (30,30) | RB | -0.0821 | -0.0147 | -0.0351 |
| | | RMSE | 0.1509 | 0.1163 | 0.1262 |
| | (100,200) | RB | -0.0289 | -0.0024 | -0.0072 |
| | | RMSE | 0.0614 | 0.0493 | 0.0524 |
| colspan="5" | Exact $\rho = 0.3246$ | | | | |
| (0, 3) | (10,10) | RB | -0.2699 | -0.0062 | -0.1819 |
| | | RMSE | 0.5648 | 0.4728 | 0.5159 |
| | (20,30) | RB | -0.1820 | -0.0028 | -0.0948 |
| | | RMSE | 0.3875 | 0.3084 | 0.3511 |
| | (30,30) | RB | -0.1694 | -0.0042 | -0.0784 |
| | | RMSE | 0.3499 | 0.2777 | 0.3193 |
| | (100,200) | RB | -0.0896 | -0.0028 | -0.0165 |
| | | RMSE | 0.1776 | 0.1276 | 0.1728 |

**Table (4).** The RB and RMSE of the estimators $\hat{\rho}_k$ and $\hat{\rho}_p$ when the data are simulated from pair normal distributions with different location and different scale parameters.

| | | | Exact $\rho = 0.9932$ | | |
|---|---|---|---|---|---|
| $(\mu_1, \sigma_1^2)$ | $(\mu_2, \sigma_2^2)$ | $(n_1, n_2)$ | | $\hat{\rho}_k$ | $\hat{\rho}_p$ |
| (0,1) | (-0.2,1.1) | (10,10) | RB | -0.1320 | -0.0618 |
| | | | RMSE | 0.1639 | 0.0923 |
| | | (20,30) | RB | -0.0636 | -0.0245 |
| | | | RMSE | 0.0778 | 0.0401 |
| | | (30,30) | RB | -0.0520 | -0.0185 |
| | | | RMSE | 0.0609 | 0.0299 |
| | | (100,200) | RB | -0.0172 | -0.0039 |
| | | | RMSE | 0.0206 | 0.0093 |
| | | | Exact $\rho = 0.6257$ | | |
| (0,1) | (2.5,4) | (10,10) | RB | −0.119 | −0.0631 |
| | | | RMSE | 0.2692 | 0.2409 |
| | | (20,30) | RB | −0.0813 | −0.0376 |
| | | | RMSE | 0.1654 | 0.1452 |
| | | (30,30) | RB | −0.0639 | −0.0263 |
| | | | RMSE | 0.1531 | 0.1361 |
| | | (100,200) | RB | −0.0293 | −0.0068 |
| | | | RMSE | 0.0624 | 0.0568 |
| | | | Exact $\rho = 0.2562$ | | |
| (0,1) | (5,2) | (10,10) | RB | −0.2552 | −0.2062 |
| | | | RMSE | 0.615 | 0.606 |
| | | (20,30) | RB | −0.1864 | −0.1051 |
| | | | RMSE | 0.4177 | 0.3978 |
| | | (30,30) | RB | −0.1616 | −0.0851 |
| | | | RMSE | 0.3886 | 0.3755 |
| | | (100,200) | RB | −0.0775 | −0.0174 |
| | | | RMSE | 0.1873 | 0.1757 |